\documentclass[twocolumn,showpacs,preprintnumbers,prd]{revtex4}
\usepackage{psfrag,graphicx,epsfig,amsmath,amssymb,bm}

\newcommand{\beq}{\begin{equation}}
\newcommand{\eeq}{\end{equation}}
\newcommand{\ord}{{\cal O}}
\newcommand{\Mkk}{M_{\rm KK}}

\begin{document}

\preprint{MZ-TH/08-33} 

\title{\boldmath
Little Randall-Sundrum models: $\epsilon_K$ strikes again
\unboldmath}

\author{M. Bauer, S. Casagrande, L. Gr\"under, U. Haisch and M. Neubert} 

\affiliation{
Institut f\"ur Physik (THEP), Johannes Gutenberg-Universit\"at,
D-55099 Mainz, Germany
}

\date{November 22, 2008}

\begin{abstract}
\noindent
A detailed phenomenological analysis of neutral kaon mixing in ``little Randall-Sundrum'' models is presented. It is shown that the constraints arising from the CP-violating quantity $\epsilon_K$ can, depending on the value of the ultra-violet cutoff, be even stronger than in the original Randall-Sundrum scenario addressing the hierarchy problem up to the Planck scale. The origin of the enhancement is explained, and a bound $\Lambda_{\rm UV}>\mbox{several $10^3$\,TeV}$ is derived, below which vast corrections to $\epsilon_K$ are generically unavoidable. Implications for non-standard $Z^0\to b\bar b$ couplings are briefly discussed.
\end{abstract}


\maketitle

\section{Introduction}
\label{sec:introduction}

Models with a warped extra dimension provide a compelling geometrical explanation of a number of mysteries left unexplained by the Standard Model (SM), most notably the hierarchy problem and the flavor puzzle. In these scenarios, first proposed by Randall and Sundrum (RS) \cite{Randall:1999ee}, one studies the SM on a background consisting of Minkowski space embedded in a slice of five-dimensional anti de-Sitter space with curvature $k$. The fifth dimension is an $S^1/Z_2$ orbifold of size $r$ labeled by a coordinate $\phi\in[-\pi,\pi]$. Two branes are located at the orbifold fixed points at $\phi=0$ (UV) and $\phi=\pi$ (IR). The metric is given by
\beq
   ds^2 = e^{-2\sigma(\phi)}\,\eta_{\mu\nu}\,dx^\mu dx^\nu 
   - r^2 d\phi^2 \,, 
\eeq
where $e^{\sigma(\phi)}$ with $\sigma(\phi)=kr|\phi|$ is called the warp factor. All SM fields except for the Higgs boson are allowed to propagate in the bulk.

While the original RS model aimed at solving the hier\-archy problem up to the Planck scale, there exist theoretical arguments in favor of lowering the ultra-violet (UV) cutoff significantly. Higher-dimensional spaces with warp factors arise naturally in flux compactifications of string theory \cite{Klebanov:2000hb, Giddings:2001yu, Kachru:2003aw,Brummer:2005sh}, and it is thus not unlikely that the RS model will have to be embedded into a more fundamental theory at some scale $\Lambda_{\rm UV}\ll M_{\rm Pl}$. From a purely phenomenological point of view, it is possible to lower the cutoff to a value only few orders of magnitude above the TeV scale, even though in this case a true solution to the hierarchy problem is postponed to higher energy. Such a scenario has been called the ``little RS'' (LRS) model \cite{Davoudiasl:2008hx}, in analogy with ``little Higgs'' models, which stabilize the Higgs mass only up to scales of order (10--100)\,TeV \cite{ArkaniHamed:2001nc,ArkaniHamed:2002qy}.

One appealing feature of such a ``volume-truncated'' RS background is that this setup allows to mitigate the strong constraints arising from electroweak precision measurements \cite{Davoudiasl:2008hx}. In particular, the $T$ parameter, which in the minimal RS scenario at tree level is given by
\beq
   T = \frac{\pi v^2}{2\cos^2\theta_W\,\Mkk^2} 
   \left( L - \frac{1}{2L} \right) ,
\eeq
can be protected from vast corrections by choosing the logarithm of
the warp factor,
\beq
   L = kr\pi
   \equiv \ln\frac{\Lambda_{\rm UV}}{\Lambda_{\rm IR}}
   \equiv -\ln\epsilon \,,
\eeq 
smaller than the RS default value $L\approx\ln(10^{16})\approx 37$, as needed to explain the hierarchy between the Planck scale and the scale of electroweak symmetry breaking. Below we refer to $L$ as the ``volume'' of the extra dimension. Taking for example $L=\ln(10^3)\approx 7$ to address the hierarchy between the weak scale and a UV scale $\Lambda_{\rm UV}\approx 10^3$\,TeV reduces the contribution to the $T$ parameter by a factor 5.4. For a light Higgs boson with mass $m_h=150$\,GeV this leads to the rather weak bound $\Mkk>1.5$\,TeV at 99\% confidence level (CL) on the ``Kaluza-Klein (KK) scale'' $\Mkk\equiv k\epsilon$, which sets the masses of the KK excitations of SM fields. Even milder constraints are obtained for a heavier Higgs boson \cite{Casagrande:2008hr}. The quoted limit, which translates into masses of the lightest KK modes of around 4\,TeV, is weaker by a factor 2.3 than the one obtained for $L\approx 37$. Compared with the original RS model, the LRS scenario thus seems to offer better prospects for a discovery of KK excitations at the Large Hadron Collider.

Since many amplitudes in the RS model are enhanced by the parameter
$L$, it is natural to ask whether lowering the UV cutoff also helps to suppress other potentially dangerous contributions to processes such as $Z^0\to b\bar b$, neutral meson mixing, and rare decays of $K$ and $B$ mesons. While values $L\ll 37$ indeed help to suppress some unwanted effects, the purpose of this article is to point out that other such contributions can even be enhanced. In particular, we show that the rather severe constraint from the CP-violating parameter $\epsilon_K$ in $K$--$\bar K$ mixing arising in RS models \cite{Csaki:2008zd,Santiago:2008vq,Blanke:2008zb,inprep} is generically even more stringent in LRS scenarios, unless the UV cutoff is raised to a value above several $10^3$\,TeV. We explain the reason for the enhancement and derive a limit on $\Lambda_{\rm UV}$ below which large corrections to $\epsilon_K$ arise generically. A comprehensive study of flavor observables in the original RS setup and the LRS model will be presented in a companion paper \cite{inprep}.

In Section~\ref{sec:mixingmatrices} we recall important results for the various flavor mixing matrices appearing in the $Z^0\to q_j\bar q_i$ couplings and the $\Delta F=1,2$ weak decay amplitudes in the RS model. We show that for certain values of the bulk mass parameters of the fermion fields these couplings develop terms that are {\em exponentially enhanced\/} in the volume factor $L$ relative to the corresponding results in the original RS model. If present, these terms give the dominant contributions to the mixing matrices, which then become largely independent of the profiles of the light quark fields. The associated flavor effects are thus UV and not infra-red (IR) dominated. This leads to a softening of the RS-GIM mechanism \cite{Agashe:2004ay,Agashe:2004cp}, which in the original RS framework is instrumental to avoid unacceptably large flavor-violating interactions. Section~\ref{sec:bound} is devoted to a systematic study of the question under which condition on the UV cutoff of the LRS model these enhanced terms arise. We find that for phenomenologically interesting values of $\Lambda_{\rm UV}$ only $K$--$\bar K$ mixing can be significantly affected. The formulae describing the CP-violating quantity $\epsilon_K$ are given in Section~\ref{sec:kaonmixing}. In Section~\ref{sec:numerics} we present a numerical analysis of $\epsilon_K$ in the LRS scenario, which confirms our general considerations. We also discuss implications of our findings for new-physics effects in other processes. Section~\ref{sec:implications} contains our conclusions.

\section{Flavor Mixing Matrices}
\label{sec:mixingmatrices}

Before going into the details of the structure of flavor-changing interactions in the RS model, we need to review some important notations and definitions from \cite{Casagrande:2008hr}, which are needed for the further discussion. The reader is referred to this reference for more details.

\subsection{Fermion profiles and Yukawa matrices}

Introducing a coordinate $t=\epsilon\,e^{\sigma(\phi)}$ along the extra dimension \cite{Grossman:1999ra}, which runs from $t=\epsilon$ on the UV brane to $t=1$ on the IR brane, we write the KK decompositions of the left-handed/right-handed components of the five-dimensional $SU(2)_L$ doublet/singlet quark fields as
\begin{eqnarray}\label{KKdecomp}
   q_L(x,t)
   &\propto& \mbox{diag}\left[ F(c_{Q_i})\,t^{c_{Q_i}} \right]
    \bm{U}_q\,q_L^{(0)}(x) + \ord\bigg( \frac{v^2}{\Mkk^2} \bigg) 
    \nonumber\\
   &&\mbox{}+ \mbox{KK excitations} \,, \nonumber\\
   q_R^c(x,t)
   &\propto& \mbox{diag}\left[ F(c_{q_i})\,t^{c_{q_i}} \right]
    \bm{W}_q\,q_R^{(0)}(x) + \ord\bigg( \frac{v^2}{\Mkk^2} \bigg) 
    \nonumber\\
   &&\mbox{}+ \mbox{KK excitations} \,, 
\end{eqnarray}
where $q=u,d$ for up- and down-type quarks, respectively. The fields are three-component vectors in flavor space. The five-dimensional fields on the left-hand side refer to interaction eigenstates, while the four-dimensional fields appearing on the right are mass eigenstates. The superscript ``(0)'' is used to denote the light SM fermions, often called ``zero modes''. In this article we are not interested in heavy KK fermions. 

The zero-mode profile \cite{Grossman:1999ra,Gherghetta:2000qt}
\beq
	F(c) = \mbox{sgn}[\cos(\pi c)]\, 
	\sqrt{ \frac{1+2c}{1-\epsilon^{1+2c}}}
\eeq
is exponentially suppressed in the volume factor $L$ if the bulk mass parameters $c_{Q_i}=+M_{Q_i}/k$ and $c_{q_i}=-M_{q_i}/k$ are smaller than the critical value $-1/2$, in which case $F(c)\sim e^{L(c+\frac12)}$. Here $M_{Q_i}$ and $M_{q_i}$ denote the masses of the five-dimensional $SU(2)_L$ doublet and singlet fermions. This mechanism explains in a natural way the large hierarchies observed in the spectrum of the quark masses \cite{Gherghetta:2000qt,Huber:2000ie}, which follow from the eigenvalues of the effective Yukawa matrices
\beq\label{Yeff}
   \bm{Y}_q^{\rm eff} = \mbox{diag}\left[ F(c_{Q_i}) \right]
   \bm{Y}_q\,\mbox{diag}\left[ F(c_{q_i}) \right]
   = \bm{U}_q\,\bm{\lambda}_q\,\bm{W}_q^\dagger \,.
\eeq 
The five-dimensional Yukawa matrices $\bm{Y}_q$ are assumed to be anarchic (i.e., non-hierarchical) with $\ord(1)$ complex elements, and $\bm{\lambda}_q$ are diagonal matrices with entries $(\lambda_q)_{ii}=\sqrt2\,m_{q_i}/v$. The unitary matrices $\bm{U}_q$ and $\bm{W}_q$ appearing in (\ref{KKdecomp}) and (\ref{Yeff}) have a hierarchical structure given by
\beq
   (U_q)_{ij} \sim
   \begin{cases} 
    \frac{\displaystyle F(c_{Q_i})}{\displaystyle F(c_{Q_j})} \,; \!\!
    & i\le j \,, \\[3mm]
    \frac{\displaystyle F(c_{Q_j})}{\displaystyle F(c_{Q_i})} \,; \!\!
    & i>j \,, 
   \end{cases} ~~
   (W_q)_{ij} \sim
   \begin{cases} 
    \frac{\displaystyle F(c_{q_i})}{\displaystyle F(c_{q_j})} \,; \!\! 
    & i\le j \,, \\[3mm]
    \frac{\displaystyle F(c_{q_j})}{\displaystyle F(c_{q_i})} \,; \!\! 
    & i>j \,. 
   \end{cases}
\eeq

Terms omitted in the first lines of the relations in (\ref{KKdecomp}) are suppressed by at least two powers of the small ratio $v/\Mkk$ and can be neglected to a good approximation. We refer to this as the ``zero-mode approximation'' (ZMA). It allows us to derive compact analytic formulae for the flavor mixing matrices that are both transparent and accurate. In our numerical analysis in Section~\ref{sec:numerics} we nevertheless employ the exact expressions for the various matrices derived in \cite{Casagrande:2008hr}.

\subsection{Gauge-boson profiles}

Whereas the profiles of the massless gluon and photon modes are flat along the extra dimension, those of the SM weak gauge bosons receive some small, $t$-dependent corrections due to electroweak symmetry breaking. Up to an overall normalization one finds \cite{Csaki:2002gy}
\beq\label{eq:chi0}
   \chi_0(t)\propto 1 - \frac{x_0^2\,t^2}{2} 
   \left( L - \frac12 + \ln t \right) + \ord(x_0^4) \,,
\eeq
where $x_0=m_{W,Z}/\Mkk$. Overlap integrals of the $t$-dependent terms with fermion profiles give rise to flavor-changing effects, which below are parameterized by matrices $\bm{\Delta}_{Q,q}^{(\prime)}$.

The profiles of the massive KK gauge bosons are given in terms of complicated combinations of Bessel functions \cite{Davoudiasl:1999tf,Pomarol:1999ad}. The lowest-lying modes live near the IR brane, i.e., in the region near $t=1$. On the other hand, the profiles of the light SM fermions are strongly peaked in the UV region near $t=\epsilon$, and for sufficiently negative values of the bulk mass parameters this region can give the dominant contributions to the relevant overlap integrals. For $t=\ord(\epsilon)$ we find
\beq
   \chi_n(t)\propto 1 - \frac{x_n^2\,t^2}{2} 
   \left( L - \frac12 + \ln t \right) + \ord(\epsilon^4) \,.
\eeq
Unlike (\ref{eq:chi0}) this approximation holds only near the UV brane.

At low energy, the exchange of virtual KK gauge bosons between SM fermions gives rise to local four-fermion interactions. The effect of summing over the infinite tower of KK bosons can be evaluated in closed form \cite{Hirn:2007bb}. Dropping irrelevant $\ord(\epsilon^2)$ constants, one finds for the sum over KK gluons or photons \cite{Casagrande:2008hr}
\beq\label{KKsum}
\begin{split}
   &\hspace{-1mm} \sum_{n\ge 1}\,\frac{\chi_n(t)\,\chi_n(t')}{m_n^2}
    = \frac{1}{4\pi\Mkk^2} \\
   &\times \left[ L\,t_<^2 - t^2 \left( \frac12 - \ln t \right)
    - t'^2 \left( \frac12 - \ln t' \right) + \frac{1}{2L} \right] ,
\end{split}
\eeq
where $t_<^2\equiv{\rm min}(t^2,t'^2)$. A similar formula holds for the sums over weak gauge bosons and their KK excitations. It will be crucial for our discussion that in this relation there is a non-factorizable term depending on both $t$ and $t'$ already at order $t^2$. Naively, from the structure of the individual gauge-boson profiles given above, one would have expected such a term to arise at order $t^2 t'^2$. This term reflects the full five-dimensional structure of the RS model, which is lost when one considers only a few low-lying KK modes. Indeed, relation (\ref{KKsum}) can also be derived by expanding the five-dimensional mixed position/momentum-space gauge-field propagator \cite{Randall:2001gb} about $q^2=0$. We will come back to the significance of this observation later.

\subsection{Flavor-changing interactions}

The non-trivial profiles of the bulk fields in the RS model lead to a plethora of flavor-changing interactions from gauge-boson exchange between fermions. For example, to order $v^2/\Mkk^2$ the couplings of the $Z^0$ boson to quarks can be parameterized as
\begin{eqnarray}
\begin{split}
   {\cal L}_{Zq\bar q} 
   &= \frac{g}{\cos\theta_W} \left[ 1
    + \frac{m_Z^2}{4\Mkk^2} \left( 1 - \frac{1}{L} \right) \right] 
    \\[1mm]
   &\quad\times Z_\mu^0 \sum_{q=u,d} \left(
    \bar q_L^{(0)} \bm{g}_L^q \gamma^\mu q_L^{(0)} 
    + \bar q_R^{(0)} \bm{g}_R^q \gamma^\mu q_R^{(0)} \right) , 
\end{split}
\end{eqnarray}
where 
\begin{eqnarray}\label{fermionZ0coupling}
\begin{split}
   \bm{g}_L^q 
   &= \left( T_3^q - \sin^2\theta_W\,Q_q \right) 
    \left[ \bm{1} - \frac{m_Z^2}{2\Mkk^2} 
    \left( L\,\bm{\Delta}_Q - \bm{\Delta}_Q' \right) \right] 
    \\[1mm]
   &\quad\mbox{}- T_3^q\,\bm{\delta}_Q \,, \\ \\ \\ \\ \\[-17mm]
   \bm{g}_R^q 
   &= - \sin^2\theta_W\,Q_q
    \left[ \bm{1} - \frac{m_Z^2}{2\Mkk^2} 
    \left( L\,\bm{\Delta}_q - \bm{\Delta}_q' \right) \right] 
    \\[1mm]
   &\quad\mbox{}+ T_3^q\,\bm{\delta}_q \,,
\end{split}
\end{eqnarray}
and $Q=U,D$ for $q=u,d$. The explicit definitions of the mixing matrices $\bm{\Delta}_A^{(\prime)}$ and $\bm{\delta}_A$ with $A=U,D,u,d$ can be found in \cite{Casagrande:2008hr}. The flavor-changing Higgs-boson couplings to quarks can be expressed in terms of the same matrices $\bm{\delta}_A$.

At low energy, exchanges of heavy bosons between SM fermions give rise to local four-fermion interactions. For $\Delta F=1$ flavor-changing neutral current (FCNC) processes (i.e., rare weak decays) these can be expressed in terms of the same flavor mixing matrices introduced above, while for $\Delta F=2$ processes (i.e., neutral meson mixing) additional mixing matrices $\widetilde{\bm{\Delta}}_A\otimes\widetilde{\bm{\Delta}}_B$ need to be introduced \cite{inprep}. For example, the phenomenologically most important contribution to $K$--$\bar K$ mixing comes from the effective interaction
\beq\label{effectiveFourQuarkVertex}
   {\cal L}_{\Delta S=2}\ni \frac{8\pi\alpha_s L}{\Mkk^2}\,
   (\widetilde{\Delta}_D)_{12}\otimes(\widetilde{\Delta}_d)_{12}\,
   (\bar d_R s_L)\,(\bar d_L s_R) \,,
\eeq
where $d\equiv d_1^{(0)}$ and $s\equiv d_2^{(0)}$.

The presence of terms proportional to the volume factor $L$ in (\ref{fermionZ0coupling}) and (\ref{effectiveFourQuarkVertex}) naively suggests that the dominant flavor-changing effects are reduced in the LRS model by a relative factor $(L_{\rm LRS}/L_{\rm RS})$ compared with the corresponding effects in the original RS model. Indeed, the alleviation of flavor constraints has been mentioned as an attractive feature of the LRS scenario \cite{Davoudiasl:2008hx}. Note, however, that in the original RS model the contributions of the $\bm{\delta}_A$ matrices in (\ref{fermionZ0coupling}) are generally of about the same magnitude as the remaining contributions \cite{Casagrande:2008hr}, and they are not reduced significantly by lowering the volume factor. We show in the present work that there is also a more subtle reason why the naive conjecture of a reduction of terms proportional to $L$ in LRS models can be flawed.

\subsection{\boldmath Mixing matrices for $\Delta F=1$ processes\unboldmath}
\label{sec:DF1}

We begin by giving explicit results, valid in the ZMA, for the flavor mixing matrices entering the $Zq_j\bar q_i$ couplings in (\ref{fermionZ0coupling}) as well as other FCNC processes. The matrices
$\bm{\Delta}^{(\prime)}_A$ are given in terms of overlap integrals of the fermion profiles in (\ref{KKdecomp}) with the $t$-dependent terms in the profile of the $Z^0$ boson in (\ref{eq:chi0}). They can be written as 
\beq\label{eq:DeltaZMA} 
  \bm{\Delta}_Q^{(\prime)}\cong \bm{U}_q^\dagger\,
   \bm{\Delta}_Q^{(\prime){\rm diag}}\,\bm{U}_q \,, \quad
  \bm{\Delta}_q^{(\prime)}\cong \bm{W}_q^\dagger\,
   \bm{\Delta}_q^{(\prime){\rm diag}}\,\bm{W}_q \,, 
\eeq 
where
\begin{widetext} 
\beq\label{eq:Deltas}
\begin{split}
  \bm{\Delta}_Q^{\rm diag} 
  = {\rm diag} \left[ \frac{1+2c_{Q_i}}{3+2c_{Q_i}}\,
   \frac{1-\epsilon^{3+2c_{Q_i}}}{1-\epsilon^{1+2c_{Q_i}}} \right] ,
   \hspace{3cm} \\[3mm]
  \bm{\Delta}_Q^{\prime\,{\rm diag}} 
  = \mbox{diag} \left[ \frac{1+2c_{Q_i}}{2(3+2c_{Q_i})^2}\,
   \frac{(5+2c_{Q_i}) (1-\epsilon^{3+2c_{Q_i}}) - 2(3+2c_{Q_i})\,L\,
         \epsilon^{3+2c_{Q_i}}}{1-\epsilon^{1+2c_{Q_i}}} \right] .
\end{split}
\eeq
\end{widetext}
Similar expressions with $c_{Q_i}$ replaced by $c_{q_i}$ hold for $\bm{\Delta}_q^{(\prime){\rm diag}}$. The symbol ``$\cong$'' in (\ref{eq:DeltaZMA}) indicates that these expressions hold in the ZMA. The corresponding exact results can be found in \cite{Casagrande:2008hr}. 

\begin{figure}[!t]
\begin{center}
\psfrag{x}[]{$c_{Q_i}$}
\psfrag{y}[b]{$(\Delta_Q^{(\prime)\,{\rm diag}})_{ii}$}
\includegraphics[height=5.0cm]{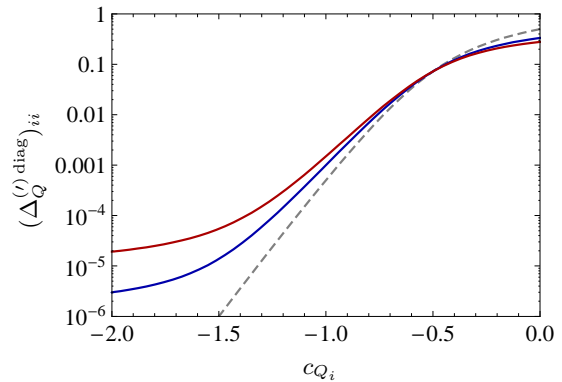}
\end{center}
\vspace{-4mm}
\caption{\label{fig:Deltas}
Elements $(\Delta_Q^{\rm diag})_{ii}$ (lower blue) and $(\Delta_Q^{\prime\,{\rm diag}})_{ii}$ (upper red) as functions of $c_{Q_i}$. The dashed line shows for comparison the zero-mode profile $F^2(c_{Q_i})/2$, which provides a good approximation to the results for $c_{Q_i}>-3/2$. We use $L=\ln(10^3)$.}
\end{figure}

The terms proportional to $\epsilon^{3+2c_{Q_i}}$ in the above formulae were omitted in \cite{Casagrande:2008hr}, because in the original RS model all $c_{Q_i}$ and $c_{q_i}$ parameters are necessarily larger than $-3/2$ in order to reproduce the correct values of the quark masses and CKM matrix elements. However, as we shall see, in LRS models we have to be more careful and keep these terms. The elements of the diagonal matrix $\bm{\Delta}_Q^{\rm diag}$ are well approximated by
\beq
   (\Delta_Q^{\rm diag})_{ii}
   \approx \left| \frac{1+2c_{Q_i}}{3+2c_{Q_i}} \right| \,
   \begin{cases}
    \, 1 \,; & \,\quad c_{Q_i} > -\frac12 \,, \\
    \, \epsilon^{-1-2c_{Q_i}} \,; & \!\!
     -\frac32 < c_{Q_i} < -\frac12 \,, \\
    \, \epsilon^2 \,; & \,\quad c_{Q_i} < -\frac32 \,,
   \end{cases} 
\eeq
and a similar relation holds for the elements of $\bm{\Delta}_Q^{\prime\,{\rm diag}}$. These formulae have the interesting feature that they exhibit an exponential dependence on the bulk mass para\-meters inside the window $-3/2<c_{Q_i}<-1/2$, while for $c_{Q_i}>-1/2$ and $c_{Q_i}<-3/2$ the dependence is far less pronounced. The region $c_{Q_i}>-1/2$ is of no concern to our discussion, since our focus is not on top-quark physics. We do not discuss it any further. Figure~\ref{fig:Deltas} shows $(\Delta_Q^{\rm diag})_{ii}$ and $(\Delta_Q^{\prime\,{\rm diag}})_{ii}$ as functions of the bulk mass parameter. The qualitatively different behavior inside and outside the window $-3/2<c_{Q_i}<-1/2$ is clearly visible. 

This behavior is readily understood from the corresponding overlap integrals. The elements of $\bm{\Delta}_Q^{\rm diag}$ are given by \cite{Casagrande:2008hr}
\beq\label{eq:Deltaoverlaps}
\begin{split}
  (\Delta_Q^{\rm diag})_{ii} 
  &= F^2(c_{Q_i}) \int_{\epsilon}^1\!dt\,t^2\,t^{2c_{Q_i}} \\ 
  &= \frac{1+2c_{Q_i}}{1-\epsilon^{1+2c_{Q_i}}} 
   \left[ \frac{t^{3+2c_{Q_i}}}{3+2c_{Q_i}} \right ]_{\epsilon}^1 \,,
\end{split}
\eeq  
and an analogous expression with an additional factor $(1/2-\ln t)$ inserted under the integral holds for the case of $\bm{\Delta}_Q^{\prime\,{\rm diag}}$. The factor $t^{2c_{Q_i}}$ in the expression above stems from the fermion profiles, see (\ref{KKdecomp}), while the weight factor $t^2$ results from the gauge-boson profile. For $c_{Q_i}>-3/2$ the overlap integrals are dominated by contributions near the IR brane at $t=1$. In this case $(\Delta_Q^{\rm diag})_{ii}\approx F^2(c_{Q_i})/(3+2c_{Q_i})$ \cite{Agashe:2003zs}, and after multiplying with the unitary matrices in (\ref{eq:DeltaZMA}) one finds that non-standard $Z^0$-boson couplings as well as other flavor-changing couplings to quarks with generation indices $i$ and $j$ are suppressed by the product $F(c_{Q_i})\,F(c_{Q_j})$ \cite{Gherghetta:2000qt}, which for $c_{Q_{i,j}}<-1/2$ is proportional to $\epsilon^{|1+c_{Q_i}+c_{Q_j}|}$, and similarly in the right-handed sector. This is the RS-GIM
mechanism \cite{Agashe:2004ay, Agashe:2004cp}. 

For $c_{Q_i}<-3/2$, on the other hand, the region near the UV brane at $t=\epsilon$ gives the dominant contribution to the integral in (\ref{eq:Deltaoverlaps}). In physical terms, this means that the overlap of the tail of the gauge-boson bulk profile with the enhanced light-quark profiles in the UV controls the size of flavor violation. We refer to this phenomenon as ``UV dominance''. The fact that $(\Delta_Q^{\rm diag})_{ii}\sim\epsilon^2$ is only weakly dependent on the bulk mass parameters in this region leads to a relaxation of the RS-GIM suppression. After multiplying with the unitary rotation matrices in (\ref{eq:DeltaZMA}), we find that flavor-changing couplings to quarks of generation $i$ and $j$ are then suppressed by $(c_{Q_i}-c_{Q_j})\,\epsilon^2 F(c_{Q_i})/F(c_{Q_j})\sim\epsilon^{2+c_{Q_j}-c_{Q_i}}$ (assuming $i<j$), and similarly in the right-handed sector. This scaling differs from the one above by a factor $\epsilon^{3+2c_{Q_j}}$. If both $c_{Q_i}$ and $c_{Q_j}$ are below the critical value $-3/2$, then this constitutes an enhancement factor $e^{L|3+2c_{Q_j}|}$ that depends exponentially on $L$.

Let us briefly comment on the mixing matrices $\bm{\delta}_A$ in (\ref{fermionZ0coupling}), which arise from the $SU(2)_L$ singlet admixture in the bulk profiles of the $SU(2)_L$ doublet SM quarks due to mixing among their KK excitations. These matrices, although often neglected in the literature, turn out to be parametrically and numerically as important as the flavor-changing effects encoded in the $\bm{\Delta}^{(\prime)}_A$ matrices \cite{Casagrande:2008hr,inprep}. The ZMA result for $\bm{\delta}_Q$ with $Q=U,D$ for $q=u,d$ reads 
\beq
   \bm{\delta}_Q\cong \bm{x}_q \bm{W}_q^\dagger\,
   \bm{\delta}_Q^{\rm diag}\,\bm{W}_q\,\bm{x}_q \,,
\eeq
where $\bm{x}_q={\rm diag}[m_{q_i}/\Mkk]$ is a diagonal matrix containing the masses of the SM quarks, and the diagonal elements $(\delta^{\rm diag}_Q)_{ii}$ can be expressed as
\beq
   (\delta_Q^{\rm diag})_{ii} 
   = \frac{1+2c_{q_i}}{1-\epsilon^{1+2c_{q_i}}} \int_{\epsilon}^1\!dt
   \left[ \frac{t^{1+c_{q_i}} - \epsilon^{1+2c_{q_i}} t^{-c_{q_i}}}%
               {1+2c_{q_i}} \right]^2 .
\eeq
An analogous expression holds in the case of $\bm{\delta}_q$ for $q=u,d$ with $c_{q_i}$ and $\bm{W}_q$ replaced by $c_{Q_i}$ and $\bm{U}_q$. Irrespective of the values of the bulk mass parameters, these overlap integrals are dominated by the region near the IR brane at $t=1$. The phenomenon of UV dominance thus never occurs for the matrices $\bm{\delta}_A$. As shown in \cite{Casagrande:2008hr}, the flavor-changing contributions mediated by these matrices are always of order $(\delta_Q)_{ij}\sim F(c_{Q_i})\,F(c_{Q_j})$ and $(\delta_q)_{ij}\sim F(c_{q_i})\,F(c_{q_j})$, i.e., they are ruled by the RS-GIM mechanism.

\subsection{\boldmath Mixing matrices for $\Delta F=2$ processes\unboldmath}
\label{sec:DF2}

Finally, we turn our attention to the mixing matrices $\widetilde{\bm{\Delta}}_A\otimes\widetilde{\bm{\Delta}}_B$ associated with $\Delta F=2$ four-quark operators, as in the effective Lagrangian (\ref{effectiveFourQuarkVertex}) \footnote{In the case of UV dominance these structures also give rise to $\Delta F=1$ contributions, which are of the same order as those from the $\bm{\Delta}_A^{(\prime)}$ matrices.}. These matrices arise from the exchange of an infinite tower of massive gauge bosons between pairs of SM fermions. Importantly, they cannot be written as simple matrix products, but only as tensor products as indicated by the $\otimes$ symbol \cite{inprep}. This fact will be crucial for our discussion. Focusing on the down-quark sector for concreteness, we obtain for $A=D$ and $B=d$ in component notation the expression
\beq\label{eq:Delta2ZMA}
\begin{split}
  & (\widetilde\Delta_D)_{ij}\otimes(\widetilde\Delta_d)_{i'j'} \\
  &\quad \cong (U_d^\dagger)_{ik}\, (U_d)_{kj}\,
   (\widetilde \Delta_{Dd})_{kl}\,  
   (W_d^\dagger)_{i'l}\,(W_d)_{lj'} \,,
\end{split}
\eeq 
where a summation over $k,l$ is understood. Here we have introduced the structures
\begin{widetext}
\beq\label{eq:Delta2s}
\begin{split}
  (\widetilde\Delta_{Dd})_{kl} 
  &= \frac12\,F^2(c_{Q_k})\,F^2(c_{d_l})
   \int_{\epsilon}^1\!dt \int_{\epsilon}^1\!dt'\,t_<^2\,\, 
   t^{2c_{Q_k}}\,(t')^{2c_{d_l}} \\
  &= \frac{(\Delta_D^{\rm diag})_{kk}}{2(1-\epsilon^{1+2c_{d_l}})}
   + \frac{(\Delta_d^{\rm diag})_{ll}}{2(1-\epsilon^{1+2c_{Q_k}})}
   - \frac{1+c_{Q_k}+c_{d_l}}{2(2+c_{Q_k}+c_{d_l})}\,
   \frac{1-\epsilon^{2(2+c_{Q_k}+c_{d_l})}}%
        {(1-\epsilon^{1+2c_{Q_k}}) (1-\epsilon^{1+2c_{d_l}})} \,,
\end{split}
\eeq
\end{widetext}
where the elements $(\Delta_{D,d}^{\rm diag})_{ii}$ have been given in the first relation in (\ref{eq:Deltas}). Analogous expressions, with obvious replacements, hold for the elements of the remaining structures $(\widetilde\Delta_{DD})_{kl}$, $(\widetilde\Delta_{dD})_{kl}$, and $(\widetilde\Delta_{dd})_{kl}$, as well as for up-type quarks. 

Assuming that all bulk mass parameters are below $-1/2$, which is always satisfied in the cases of interest, the elements $(\widetilde \Delta_{Dd})_{kl}$ are well approximated by
\begin{widetext}
\beq\label{eq:Delta2approx} 
   (\widetilde \Delta_{Dd})_{kl}\approx 
   \frac{1}{2(2+c_{Q_k}+c_{d_l})} \, 
   \begin{cases} 
    \, \displaystyle \frac{(1+2c_{Q_k}) (1+2c_{d_l})
     (3+c_{Q_k}+c_{d_l})}{(3+2c_{Q_k}) (3+2c_{d_l})}\,
     \epsilon^{-2-2c_{Q_k}-2c_{d_l}} \,;
    & c_{Q_k} + c_{d_l} > -2 \,, \\[3mm]
    \, (1+c_{Q_k}+c_{d_l})\,\epsilon^2 \,; 
    & c_{Q_k} + c_{d_l} < -2 \,.
   \end{cases} 
\eeq
\end{widetext}
In the first case $(\widetilde\Delta_{Dd})_{kl}\sim F^2(c_{Q_k})\,F^2(c_{d_l})$, and after taking into account the effect of the unitary rotation matrices in (\ref{eq:Delta2ZMA}) we find that $(\widetilde\Delta_D)_{ij}\otimes(\widetilde\Delta_d)_{i'j'}\sim F(c_{Q_i})\,F(c_{Q_j})\,F(c_{d_{i'}})\,F(c_{d_{j'}})$, in accordance with the RS-GIM mechanism. In the second case, the dominant contribution arises from the $\epsilon^{2(2+c_{Q_k}+c_{d_l})}$ piece in the numerator of the last term in (\ref{eq:Delta2s}), which is an effect that results from the UV region of the double parameter integral, where both $t$ and $t'$ are of order $\epsilon$. This term arises because for $c_{Q_k}+c_{d_l}<-2$ the weight factor $t^2_<$ appearing in the overlap integral (\ref{eq:Delta2s}) does not fall off sufficiently fast near the UV brane to compensate the strong increase of the fermion profiles. Note that if the non-factorizable term in (\ref{KKsum}) would be of the naively expected form $t^2 t'^2$, then UV dominance would only set in under the much stronger condition $c_{Q_k}+c_{d_l}<-3$ and would produce a contribution $(\widetilde \Delta_{Dd})_{kl}\sim\epsilon^4=e^{-4L}$, which would be entirely negligible for all realistic values of the volume factor $L$. In order to obtain the larger effect exhibited in (\ref{eq:Delta2approx}) it is essential to sum over the infinite KK tower, thereby probing the full five-dimensional structure of the RS background. 

If the critical condition $c_{Q_k}+c_{d_l}<-2$ is satisfied, then the elements $(\widetilde\Delta_{Dd})_{kl}$ are only weakly dependent on the bulk mass parameters. The corresponding flavor effects are thus not RS-GIM suppressed. Taking into account the effect of the unitary rotation matrices in (\ref{eq:Delta2ZMA}), we find that the tensor structures relevant to neutral meson mixing in the down-quark sector scale as $(\widetilde\Delta_D)_{ij}\otimes(\widetilde\Delta_d)_{ij}\sim(c_{Q_i}-c_{Q_j})\,(c_{d_i}-c_{d_j})\,\epsilon^2/[F^2(c_{Q_j})\,F^2(c_{d_j})]$ (assuming $i<j$). If and only if $c_{Q_j}+c_{q'_j}<-2$ for the {\em larger\/} of the two flavor indices, then this contribution is enhanced compared with the corresponding contribution in the RS case by a factor $e^{2L|2+c_{Q_j}+c_{d_j}|}\gg 1$.

\section{Bounds on the UV cutoff}
\label{sec:bound}

We now analyze the implications of the observations made in the previous section for the phenomenology of flavor-changing processes in the LRS model and investigate which processes are susceptible to the phenomenon of UV dominance. Instead of the original bulk mass parameters $c_{A_i}$ with $A=Q,u,d$, it is useful to introduce new parameters $d_{A_i}\equiv\mbox{max}(-c_{A_i}-1/2, 0)$, which measure the distance to the critical value $-1/2$. If $c_{A_i}>-1/2$ then $d_{A_i}=0$. With this convention, the following scaling relations hold for the quark masses and the Wolfenstein parameters $\lambda$ and $A$ \cite{Csaki:2008zd,Casagrande:2008hr,Huber:2003tu}:
\beq\label{eq:scalings}
\begin{split}
   \frac{\sqrt2\,m_{q_i}}{v} 
   &\sim |Y|\,e^{-L(d_{Q_i}+d_{q_i})} \,, \\[-0.4mm]
   \lambda &\sim e^{-L(d_{Q_1}-d_{Q_2})} \,, \\[0.6mm]
   A &\sim e^{-L(3d_{Q_2}-2d_{Q_1}-d_{Q_3})} \,,
\end{split} 
\eeq
where $Y$ represents a dimensionless Yukawa coupling, and we have omitted $\ord(1)$ factors. In the LRS model, the logarithm of the original RS warp factor $L_{\rm RS}=\ln(M_{\rm Pl}/\Lambda_{\rm IR})\approx 37$ is replaced by a much smaller factor $L_{\rm LRS}=\ln(\Lambda_{\rm UV}/\Lambda_{\rm IR})$ with $\Lambda_{\rm UV}\approx 10^3$\,TeV or a value of similar magnitude. In order to still satisfy the conditions for the quark masses and Wolfenstein parameters, the $d_{A_i}$ coefficients of the LRS model must be related to those of the original RS model by
\beq
   d_{A_i}^{\rm LRS} 
   \approx\frac{L_{\rm RS}}{L_{\rm LRS}}\,d_{A_i}^{\rm RS} \,.
\eeq

Assuming anarchic Yukawa couplings, $|Y|=\ord(1)$, it follows from (\ref{eq:scalings}) that
\beq\label{eq:LRrelations} 
\begin{split}
   L\,(d_{Q_1}+d_{d_1}) 
   &\sim \ln\left( \frac{|Y|\,v}{\sqrt2\,m_d} \right)
    \approx 11.0 \,, \\
   L\,(d_{Q_2}+d_{d_2})
   &\sim \ln\left( \frac{|Y|\,v}{\sqrt2\,m_s} \right)
    \approx 8.2 \,, \\
   L\,(d_{Q_3}+d_{d_3})
   &\sim \ln\left( \frac{|Y|\,v}{\sqrt2\,m_b} \right)
    \approx 4.4 \,.
\end{split}
\eeq
On the other hand, from the relations
\beq
\begin{split}
   \frac{m_t}{m_d} 
   &\sim e^{L(d_{Q_1}-d_{Q_3}+d_{d_1}-d_{u_3})}
    \sim \frac{1}{A\lambda^3}\,e^{L d_{d_1}} \,, \\
   \frac{m_t}{m_s} 
   &\sim e^{L(d_{Q_2}-d_{Q_3}+d_{d_2}-d_{u_3})}
    \sim \frac{1}{A\lambda^2}\,e^{L d_{d_2}} \,, \\ 
   \frac{m_t}{m_b} 
   &\sim e^{L(d_{d_3}-d_{u_3})} \sim e^{L d_{d_3}} \,,
\end{split}
\eeq
in which we have used the fact that the right-handed top quark is localized near the IR brane and hence $d_{u_3}=0$, it follows that
\beq\label{eq:Rrelations}
\begin{split}
   L\,d_{d_1} 
   &\sim \ln\left( A\lambda^3\,\frac{m_t}{m_d} \right)
    \approx 6.1 \,, \\
   L\,d_{d_2} 
   &\sim \ln\left( A\lambda^2\,\frac{m_t}{m_s} \right)
    \approx 4.8 \,, \\
   L\,d_{d_3} 
   &\sim \ln\left( \frac{m_t}{m_b} \right) \approx 4.2 \,.
\end{split}
\eeq
Combining (\ref{eq:LRrelations}) and (\ref{eq:Rrelations}) we obtain
\beq\label{eq:Lrelations}
\begin{split}
   L\,d_{Q_1} 
   &\sim \ln\left( \frac{1}{A\lambda^3}\,\frac{|Y|\,v}{\sqrt2\,m_t}
    \right) \approx 4.9 \,, \\
   L\,d_{Q_2} 
   &\sim \ln\left( \frac{1}{A\lambda^2}\,\frac{|Y|\,v}{\sqrt2\,m_t}
    \right) \approx 3.4 \,, \\
   L\,d_{Q_3} 
   &\sim \ln\left( \frac{|Y|\,v}{\sqrt2\,m_t} \right) \approx 0.2 \,.
\end{split}
\eeq
The numerical values quoted in these relations correspond to the input parameters used in \cite{Casagrande:2008hr}.

In Section~\ref{sec:DF1} we saw that the elements $(\Delta_A^{(\prime)})_{ij}$ with $A=Q,q$ and $i<j$ of the flavor mixing matrices parameterizing $\Delta F=1$ processes become UV-dominated and thus enhanced with respect to the original RS case if $d_{A_j}>1$, corresponding to $c_{A_j}<-3/2$. Note that this is a condition on the bulk mass parameter of the heavier quark in the transition, so $j\ge 2$ and the bulk mass parameters of the first-generation quarks are irrelevant in this context. From (\ref{eq:Rrelations}) and (\ref{eq:Lrelations}) it is clear that the strongest bound in the down-quark sector follows from the relation $L\,d_{d_2}\sim 4.8$. In order to avoid a large enhancement of rare kaon decays in the LRS model relative to the RS model, we must require that $d_{d_2}^{\rm LRS}<1$, which yields the condition
\beq\label{weakbound}
   L_{\rm LRS} > 4.8 ~~\Rightarrow~~
   \frac{\Lambda_{\rm UV}}{\Lambda_{\rm IR}} > 120 \,.
   \quad (\Delta S=1)
\eeq
All other relations, including those involving up-type quarks, lead to weaker limits on $L_{\rm LRS}$. Barring a conspiracy of undetermined $\ord (1)$ factors, $\Delta F=1$ transitions are thus not subject to UV dominance provided the UV cutoff is chosen in the few hundred TeV range or higher. Likewise, the $Z^0\to b\bar b$ couplings are not affected by UV dominance as long as the bound (\ref{weakbound}) is satisfied.

In the realm of $\Delta F=2$ processes, $K$--$\bar K$ mixing is most likely to be affected by UV dominance in the flavor sector. In Section~\ref{sec:DF2} we have shown that the relevant condition is $d_{A_2}+d_{B_2}>1$, corresponding to $c_{A_2}+c_{B_2}<-2$ with $A,B=D,d$. The discussion in the following section will reveal that phenomenologically most relevant by far is the mixed-chirality case. In order to avoid UV dominance in this channel we must require that $d_{Q_2}+d_{d_2}<1$, which translates into the bound
\beq\label{ourbound}
   L_{\rm LRS} > 8.2 ~~\Rightarrow~~
   \frac{\Lambda_{\rm UV}}{\Lambda_{\rm IR}} > 3600 \,.
   \quad (\Delta S=2)
\eeq
Four-quark operators containing only right-handed quarks receive an even stronger enhancement, which is present as long as $\Lambda_{\rm UV}/\Lambda_{\rm IR}<15\cdot 10^3$, however their contribution to the observables in $K$--$\bar K$ mixing are in any case strongly suppressed. To the extent that the RS contribution to the CP-violating quantity $\epsilon_K$ is dominated by the mixed-chirality operator in (\ref{effectiveFourQuarkVertex}), we find that the ratio of the new-physics contributions obtained in the LRS and RS scenarios scales like
\beq\label{eq:Lbehavior}
\begin{split}
   \frac{|\Delta\epsilon_K|_{\rm LRS}}{|\Delta\epsilon_K|_{\rm RS}}
   &\approx \frac{L_{\rm LRS}}{L_{\rm RS}}\,\,\mbox{max}\left[ 1,\,
    e^{-2L_{\rm LRS}} \left( \frac{|Y|\,v}{\sqrt2\,m_s} \right)^2
    \right] \\
   &\approx \frac{L_{\rm LRS}}{37}\,\,\mbox{max}\left[ 1,\,
    e^{2(8.2-L_{\rm LRS})} \right] .
\end{split}
\eeq
This ratio becomes minimal for $L_{\rm LRS}\approx 8.2$ and exponentially increases for smaller values of the volume factor. It follows that $K$--$\bar K$ mixing is affected by UV dominance in a large class of LRS models. One of the original motivations of the LRS scenario, namely to mitigate the severe constraints from flavor physics on the value of the KK scale $\Mkk$, becomes invalidated when the UV cutoff is taken below a value of several thousand TeV. This conclusion will be supported by the detailed numerical analysis in Section~\ref{sec:numerics}.

\section{Neutral kaon mixing}
\label{sec:kaonmixing}

We adopt the following general parametrization of new-physics effects
in $K$--$\bar K$ mixing \cite{Ciuchini:1998ix}:
\beq\label{eq:Heff}
   {\cal H}_{\rm eff}^{\Delta S=2} 
   = \sum_{i=1}^5 C_i\,Q_i + \sum_{i=1}^3 \tilde C_i\,\tilde Q_i \,.
\eeq
The operators relevant to our discussion are
\beq
\begin{aligned}
   Q_1 &= (\bar d_L\gamma^\mu s_L)\,
    (\bar d_L\gamma_\mu s_L) \,, & 
   \tilde Q_1 & = (\bar d_R\gamma^\mu s_R)\,
    (\bar d_R\gamma_\mu s_R) \,, \hspace{-0.25cm} \\[1mm]
   Q_4 &= (\bar d_R s_L)\,(\bar d_L s_R) \,, &
   Q_5 &= (\bar d_R^\alpha s_L^\beta)\,(\bar d_L^\beta s_R^\alpha) \,,
\end{aligned}
\eeq
where a summation over color indices $\alpha,\beta$ is understood. We write the Wilson coefficients as a sum of a SM and a new-physics contribution, $C_i\equiv C_i^{\rm SM}+C_i^{\rm RS}$, where in the SM only $C_1^{\rm SM}$ is non-zero. Using the results of \cite{inprep}, we obtain for the contributions arising in the minimal RS model
\beq\label{eq:Csd}
\begin{split}
   C_1^{\rm RS} &= \frac{4\pi L}{\Mkk^2}\, 
    (\widetilde\Delta_D)_{12}\otimes(\widetilde\Delta_D)_{12} 
    \left( \frac{\alpha_s}{3} + 1.04\,\alpha \right) , \\
   \tilde C_1^{\rm RS} &= \frac{4\pi L}{\Mkk^2}\, 
    (\widetilde\Delta_d)_{12}\otimes(\widetilde\Delta_d)_{12} 
   \left( \frac{\alpha_s}{3} + 0.15\,\alpha \right) , \\
   C_4^{\rm RS} &= \frac{4\pi L}{\Mkk^2}\, 
    (\widetilde\Delta_D)_{12}\otimes(\widetilde\Delta_d)_{12} 
    \left( - 2\alpha_s \right) , \\
   C_5^{\rm RS} &= \frac{4\pi L}{\Mkk^2}\, 
    (\widetilde\Delta_D)_{12}\otimes(\widetilde\Delta_d)_{12} 
    \left( \frac{2\alpha_s}{3} + 0.30\,\alpha \right) ,
\end{split}
\eeq 
while $C_{2,3}^{\rm RS}=0$ and $\tilde C_{2,3}^{\rm RS}=0$. These matching conditions hold at a scale $\mu=\ord(\Mkk)$. The expressions in brackets refer, in an obvious way, to the contributions from KK gluons and massive electroweak gauge bosons. In the latter contributions we have set $\sin^2\theta_W=0.25$ corresponding to a matching scale $\mu\approx 3$\,TeV to derive the numerical coefficients in front of the $\ord(\alpha)$ terms. Contributions from flavor-changing Higgs-boson exchange are of order $v^4/\Mkk^4$ and therefore have been neglected in (\ref{eq:Csd}). Notice that even in the case of $C_1^{\rm RS}$ the QCD contribution is larger by a factor of about 3 than the combined QED and electroweak effects. KK gluon exchange thus dominates the mixing amplitudes for all neutral mesons. In the particular case of $K$--$\bar K$ mixing, the contribution of $C_4^{\rm RS}$ receives such a strong enhancement (see below) that all other contributions can be neglected. It follows that the mixing amplitudes are not very sensitive to the choice of the electroweak gauge group of the model. Hence, our analysis not only applies to the original RS model with the SM gauge group in the bulk, but also to models with an extended gauge sector and custodial symmetry \cite{Agashe:2003zs}.

The tree-level expressions for the Wilson coefficients given in
(\ref{eq:Csd}) must be evolved, using the renormalization group, down to a scale $\mu\approx 2$\,GeV, at which the hadronic matrix elements of the four-quark operators can be evaluated using lattice QCD. This is accomplished with the help of formulae compiled in the literature \cite{Ciuchini:1998ix}. The hadronic matrix elements of the various operators are customarily expressed in terms of parameters $B_i$. For the operators relevant to our analysis, one has
\beq
\begin{split}
   \langle K^0|\,Q_1\,|\bar K^0\rangle 
   &= \langle K^0|\,\tilde Q_1\,|\bar K^0\rangle 
    = \frac{m_K f_K^2}{3}\,B_1 \,, \\
   \langle K^0|\,Q_4\,|\bar K^0\rangle 
   &= \left( \frac{m_K}{m_s+m_d} \right)^2 
    \frac{m_K f_K^2}{4}\,B_4 \,, \\
   \langle K^0|\,Q_5\,|\bar K^0\rangle 
   &= \left( \frac{m_K}{m_s+m_d} \right)^2
    \frac{m_K f_K^2}{12}\,B_5 \,,
\end{split}
\eeq
where $B_i\equiv B_i(\mu)$ and $m_{d,s}\equiv m_{d,s}(\mu)$ evaluated in the $\overline{\rm MS}$ scheme. In our numerical analysis we employ the kaon $B_i$ parameters from \cite{Lubicz:2008am}. Furthermore, we use $m_K=497.6$\,MeV and $f_K=(156.1\pm 0.8)$\,MeV for the kaon mass and decay constant. 

In terms of the effective Hamiltonian (\ref{eq:Heff}), the CP-violating quantity $\epsilon_K$ is given by
\beq
   \epsilon_K 
   = \frac{\kappa_\epsilon\,e^{i\varphi_\epsilon}}%
          {\sqrt2\,(\Delta m_K)_{\rm exp}}\, 
    \mbox{Im}\,\langle K^0|\,{\cal H}_{\rm eff}^{\Delta S=2}\, 
    |\bar K^0\rangle \,,
\eeq
where $\varphi_\epsilon=(43.51\pm 0.05)^\circ$, and $(\Delta m_K)_{\rm exp}=(3.483\pm 0.006)\cdot 10^{-13}$\,MeV is the experimental value of the $K_L$--$K_S$ mass difference \cite{Amsler:2008zz}. The suppression factor $\kappa_{\epsilon}=0.92\pm 0.02$ \cite{Buras:2008nn} parameterizes the effects due to the imaginary part of the isospin-zero amplitude in $K\to\pi\pi$ decays \cite{Anikeev:2001rk,Andriyash:2003ym}. 

Taking into account all indirect constraints from the global unitarity-triangle fit \cite{Charles:2004jd}, we obtain the SM prediction $|\epsilon_K|_{\rm SM}=(1.8\pm 0.5)\cdot 10^{-3}$, where the quoted error corresponds to a frequentist 68\% CL. The dominant theoretical uncertainty arises from the lattice QCD prediction $B_1(2\,{\rm GeV)}=0.55\pm 0.05$ \cite{Lubicz:2008am}, which has been scanned in its range to obtain the best fit value. Within errors the SM prediction agrees with the experimental value $|\epsilon_K|_{\rm exp}=(2.229\pm 0.010)\cdot 10^{-3}$ \cite{Amsler:2008zz}. We do not attempt a prediction for $\Delta m_K$, which is plagued by very large hadronic uncertainties.

\section{Numerical analysis}
\label{sec:numerics}

It is useful for many considerations to have a default set of input parameters, which is consistent with the experimental values of the quark masses and CKM parameters. We use $\Mkk=3$\,TeV as the default KK scale and set $L_{\rm LRS}=\ln(10^{3})\approx 7$ and $L_{\rm RS}=\ln(10^{16})\approx 37$ for the logarithm of the warp factor, unless indicated otherwise. For the five-dimensional Yukawa matrices we take 
\beq
\begin{split}
   \bm{Y}_u &= \left( 
    \begin{array}{lll}   
      0.59 \, e^{-i \, 33^\circ} & 
      1.26 \, e^{i \, 24^\circ} &  
      0.69 \, e^{-i \, 21^\circ} \\
      0.94 \, e^{i \, 170^\circ} & 
      2.92 \, e^{-i \, 168^\circ} &  
      1.66 \, e^{i \, 31^\circ} \\
      2.59 \, e^{-i \, 108^\circ} & 
      2.75 \, e^{-i \, 85^\circ} &  
      2.69 \, e^{i \, 170^\circ} 
    \end{array}
   \right) , \\[2mm] 
   \bm{Y}_d &= \left( 
    \begin{array}{lll}
      2.45 \, e^{i \, 111^\circ} & 
      0.13 \, e^{i \, 122^\circ} &  
      0.96 \, e^{-i \, 120^\circ} \\
      1.07 \, e^{i \, 154^\circ} & 
      2.31 \, e^{-i \, 84^\circ} &  
      0.92 \, e^{-i \, 45^\circ} \\
      2.85 \, e^{i \, 146^\circ} & 
      0.84 \, e^{-i \, 139^\circ} &  
      2.48 \, e^{i \, 72^\circ}  
    \end{array}
   \right) . 
\end{split}
\eeq
Our default set of bulk mass parameters is
\begin{eqnarray}
\begin{aligned}
  c_{Q_1}^{\rm LRS} &= -1.10 \,, & \ 
  c_{Q_2}^{\rm LRS} &= -0.94 \,, & \
  c_{Q_3}^{\rm LRS} &= -0.54 \,, \\
  c_{u_1}^{\rm LRS} &= -1.69 \,, & \ 
  c_{u_2}^{\rm LRS} &= -1.01 \,, & \
  c_{u_3}^{\rm LRS} &= -0.09 \,, \\
  c_{d_1}^{\rm LRS} &= -1.74 \,, & \ 
  c_{d_2}^{\rm LRS} &= -1.42 \,, & \ 
  c_{d_3}^{\rm LRS} &= -1.12 \,, \hspace{6mm}
\end{aligned}
\end{eqnarray}
which corresponds to 
\beq
\begin{aligned}
  c_{Q_1}^{\rm RS} &= -0.59 \,, & \ 
  c_{Q_2}^{\rm RS} &= -0.55 \,, & \
  c_{Q_3}^{\rm RS} &= -0.45 \,, \\
  c_{u_1}^{\rm RS} &= -0.70 \,, & \ 
  c_{u_2}^{\rm RS} &= -0.57 \,, & \
  c_{u_3}^{\rm RS} &= -0.09 \,, \\
  c_{d_1}^{\rm RS} &= -0.71 \,, & \ 
  c_{d_2}^{\rm RS} &= -0.65 \,, & \
  c_{d_3}^{\rm RS} &= -0.59 \,,
\end{aligned}
\eeq 
in the original RS setup. The Yukawa matrices have been obtained by random choice, subject to the constraints $|(Y_q)_{ij}|<3$. The bulk mass parameters have then been adjusted such that the observed quark masses and CKM parameters are reproduced within 68\% CL.

We begin our analysis with a discussion of the values of the Wilson coefficients in (\ref{eq:Csd}), which enter the prediction for $\epsilon_K$. For the default input parameters we find in the LRS model
\beq\label{eq:CLRS}
\begin{split}
   C_1^{\rm LRS} &= (1.15 - 1.91\hspace{0.5mm} i)\cdot 10^{-15} \,, 
    \\[1mm]
   \widetilde C_1^{\rm LRS} &= (1.51 + 0.17\hspace{0.5mm} i)
    \cdot 10^{-16} \,, \\[1mm]
   C_4^{\rm LRS} &= (-1.56 + 0.63\hspace{0.5mm} i)\cdot 10^{-15} \,,
    \\[1mm]
   C_5^{\rm LRS} &= (5.45 - 2.18\hspace{0.5mm} i) \cdot 10^{-16} \,.
\end{split}
\eeq
Here and in the following the quoted numbers are obtained after adjusting the phases of the SM quark fields according to the standard CKM phase convention. The corresponding RS values are 
\beq\label{eq:CRS}
\begin{split}
   C_1^{\rm RS} &= (-3.03 - 1.33\hspace{0.5mm} i)\cdot 10^{-16} \,,
    \\[1mm]
   \widetilde C_1^{\rm RS} &= (6.05 + 3.87\hspace{0.5mm} i)
    \cdot 10^{-20} \,, \\[1mm]
   C_4^{\rm RS} &= (-1.28 + 2.35\hspace{0.5mm} i) \cdot 10^{-17} \,,
    \\[1mm]
   C_5^{\rm RS} &= (4.46 - 8.19\hspace{0.5mm} i) \cdot 10^{-18} \,.
\end{split}
\eeq
Focusing on the magnitude of the imaginary parts of the Wilson coefficients, we see that the LRS contributions to the operators $Q_1$, $\tilde Q_1$, and $Q_{4,5}$ are enhanced compared to the RS contributions by factors of 14, 439, and 27, respectively. These enhancements follow the pattern discussed in Section~\ref{sec:bound}.

The implications of the results (\ref{eq:CLRS}) and (\ref{eq:CRS}) become clear once we understand how the Wilson coefficients enter the
prediction for $\epsilon_K$. The approximation
\beq
   |\Delta\epsilon_K|_{\rm RS} 
   \propto {\rm Im}\!\left[ C_1^{\rm RS} + \widetilde C_1^{\rm RS} 
   + 115 \left( C_4^{\rm RS} + \frac{C_5^{\rm RS}}{3} \right)
   \right] ,
\eeq
valid for a matching scale $\mu\approx 3$\,TeV, implies that, barring a conspiracy of parameters, the operator $Q_4$ gives the dominant contribution to $|\Delta\epsilon_K|$. Indeed, for our default parameters the contribution of $C_4^{\rm LRS}$ ($C_4^{\rm RS}$) to $|\Delta\epsilon_K|$ is larger than that of $C_1^{\rm LRS}$ ($C_1^{\rm RS}$) by a factor of about 40 (20). Using our default parameter set, we find for the new-physics contribution to $\epsilon_K$ in the LRS model $|\Delta\epsilon_K|_{\rm LRS}\approx 22\cdot 10^{-3}$, which exceeds the value $|\Delta\epsilon_K|_{\rm RS}\approx 0.8\cdot 10^{-3}$ obtained in the RS scenario by a factor of more than 25. Notice that the former value is incompatible with the experimental determination of $|\epsilon_K|$, while the latter one is consistent with it at 95\% CL. 

\begin{figure}[!t]
\begin{center}
\vspace{3mm}
\psfrag{z}[b]{$\Lambda_{\rm UV}/\Lambda_{\rm IR}$}
\psfrag{x}[]{$L$}
\psfrag{y}[b]{$\displaystyle\frac{|\Delta\epsilon_K|_{\rm LRS}}{|\Delta\epsilon_K|_{\rm RS}}$}
\includegraphics[width=7.0cm]{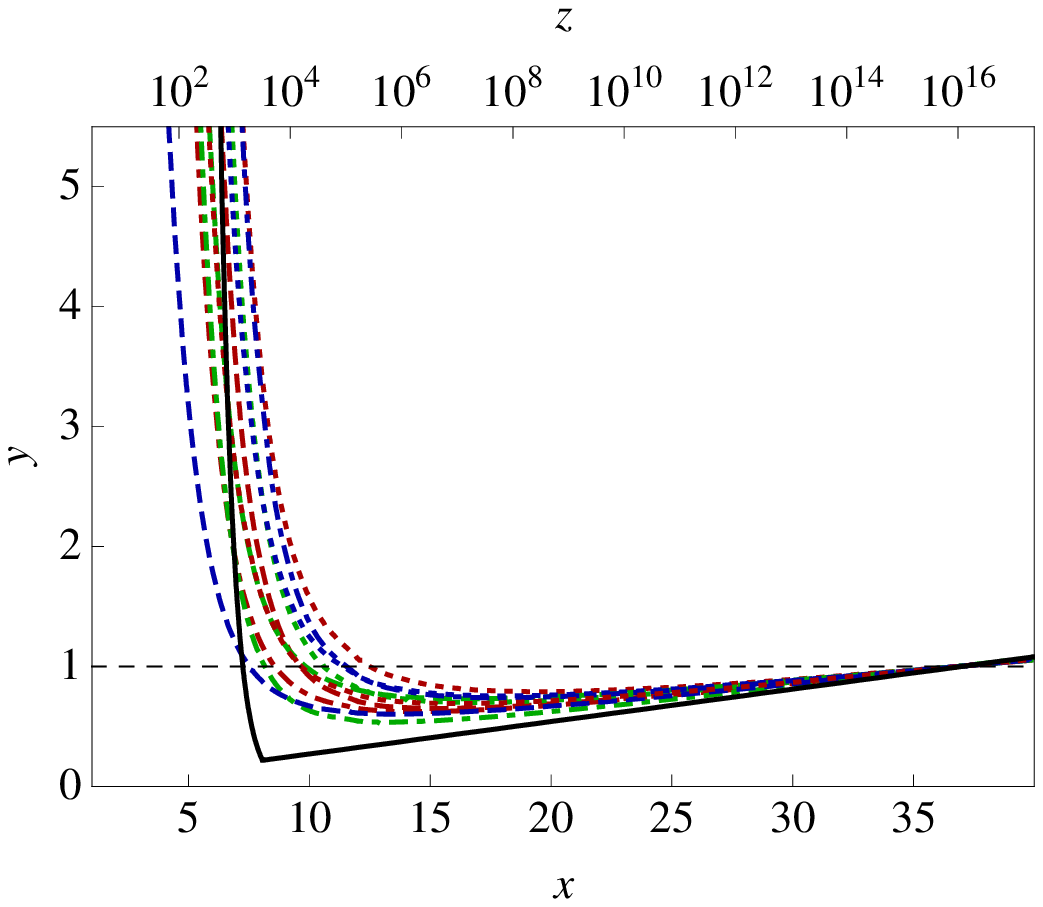}

\vspace{4mm}
\includegraphics[width=7.0cm]{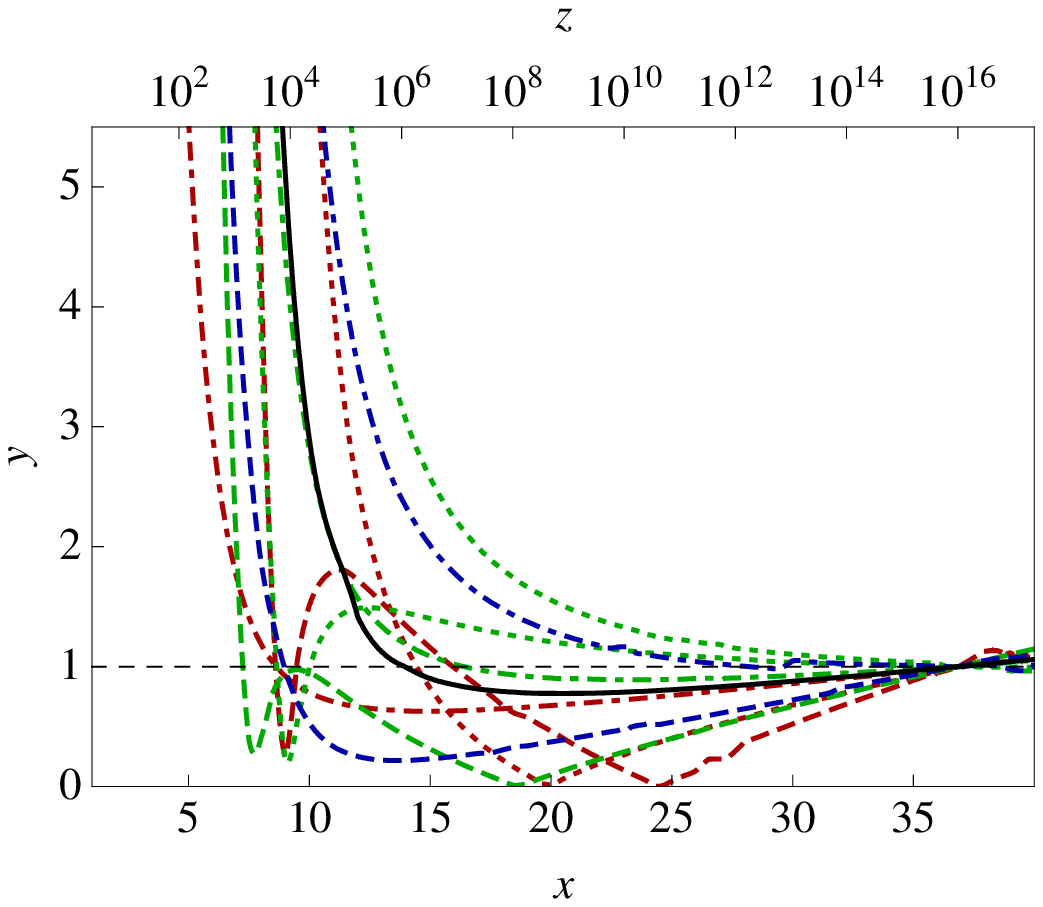}
\end{center}
\vspace{-4mm}
\caption{\label{fig:epsK} 
Ratio $|\Delta\epsilon_K|_{\rm LRS}/|\Delta\epsilon_K|_{\rm RS}$ as a function of the volume factor. The shown lines correspond to different sets of input parameters that lead to the correct quark masses and CKM parameters. While in the upper panel a random sample of predictions is shown, the lower panel contains only curves for which $|\Delta\epsilon_K|_{\rm RS}<1.0\cdot 10^{-3}$. See text for details.}
\end{figure}

In Figure~\ref{fig:epsK} we show the ratio $|\Delta\epsilon_K|_{\rm LRS}/|\Delta\epsilon_K|_{\rm RS}$ as a function of the volume factor for twenty different sets of input parameters. In the upper panel a random sample of predictions is shown, while the lower panel features only curves that satisfy $|\Delta\epsilon_K|_{\rm RS}<1.0\cdot 10^{-3}$ for fixed $\Mkk=3$\,TeV. The latter constraint guarantees that the theoretical prediction for $|\epsilon_K|$ obtained by adding the SM and RS contributions is consistent with the experimental value at 95\% CL. The difference between the two plots is not difficult to understand. The curves in the upper panel correspond, without exception, to the generic case where $Q_4$ gives the by far largest contribution to $\epsilon_K$, typically a factor 10 or more bigger than the SM prediction \cite{Csaki:2008zd,Santiago:2008vq,Blanke:2008zb,inprep}. In consequence, the $L$ dependence of the curves is given approximately by (\ref{eq:Lbehavior}), as indicated by the solid line in the upper plot. In order to achieve an acceptable amount of CP violation in the kaon sector, $|\Delta\epsilon_K|_{\rm RS}<1.0\cdot 10^{-3}$, the contribution arising from $Q_4$ has to be suppressed by a suitable adjustment of parameters. The most obvious way to reduce the magnitude of $C_4^{\rm RS}$ consists in increasing $|(Y_d)_{ij}|$ or tuning the phase of the associated combination of Yukawa couplings. As a result, the $L$ dependence of $|\Delta\epsilon_K|_{\rm LRS}/|\Delta\epsilon_K|_{\rm RS}$ tends to become more complicated. This is illustrated in the lower panel of Figure~\ref{fig:epsK}. The solid line in this plot corresponds to the result obtained for our default parameter set. While the exact shape of the curves is determined by a complicated interplay of the different contributions to $\epsilon_K$, mainly those arising from $Q_1$ and $Q_4$, we see that all curves still share the feature that below a certain value of $L$ they start to rise exponentially, signaling the onset of UV dominance. Beyond this point the Wilson coefficient of $Q_4$ always gives the dominant contribution to $\epsilon_K$. The onset of UV dominance occurs for $\Lambda_{\rm UV}/\Lambda_{\rm IR}$ in the ballpark of a few $10^3$ or higher, which agrees with the generic bound derived in (\ref{ourbound}).

We now turn to a brief discussion of non-standard effects in the $Z^0\to b\bar b$ couplings. As explained in Section~\ref{sec:bound}, in this case UV dominance does not occur for realistic values of the volume factor. The $L$-dependent terms in the relevant couplings in (\ref{fermionZ0coupling}) are then indeed strongly reduced in the LRS model compared with the RS case. However, as remarked earlier, a similar reduction is not operational for the terms parameterized by the $\bm{\delta}_{D,d}$ matrices. For our default parameter set, we obtain the following corrections $\Delta g_{L,R}^b\equiv (g_{L,R}^d)_{33}-(g_{L,R}^b)_{\rm SM}$ in the LRS model:
\beq\label{eq:LRZLRS}
\begin{split}
   (\Delta g^b_L)_{\rm LRS} &= (-0.9-7.1)\cdot 10^{-4}
    = -8.0\cdot 10^{-4} \,, \\[1mm]
   (\Delta g^b_R)_{\rm LRS} &= (-0.1+1.3)\cdot 10^{-6} 
    = +1.2\cdot 10^{-6} \,.
\end{split}
\eeq
The two numbers quoted in the first step refer to the contributions stemming from the $(\Delta_{D,d}^{(\prime)})_{33}$ and $(\delta_{D,d})_{33}$ terms, respectively. In the original RS model we find instead 
\beq\label{eq:LRZRS}
\begin{split}
   (\Delta g^b_L)_{\rm RS} &= (-5.1-11.8)\cdot 10^{-4}
    = -16.9 \cdot 10^{-4} \,, \\[1mm]
   (\Delta g^b_R)_{\rm RS} &= (-0.2+\phantom{1}1.5)\cdot 10^{-6} 
    = +\phantom{1}1.3\cdot 10^{-6} \,.
\end{split}
\eeq
Comparing the numbers in (\ref{eq:LRZLRS}) with those in (\ref{eq:LRZRS}), we see that the left-handed coupling is larger than the right-handed one by more than a factor 600 (1200) in the LRS (RS)
case. More importantly, the left-handed coupling in the LRS scenario
is suppressed with respect to the one in the original RS setup by a
factor of about 2, while in the case of the right-handed coupling the
ratio between the LRS and RS result is only slightly smaller than~1. The suppression of the left-handed coupling is understood by noting that the term proportional to $(\Delta_D)_{33}$ in the bracket appearing in the first line of (\ref{fermionZ0coupling}) is reduced by a factor of $L_{\rm RS}/L_{\rm LRS}\approx 5$, whereas $(\delta_D)_{33}$ is only weakly dependent on the volume factor through the bulk mass parameters. This pattern of suppressions, combined with the fact that for our reference point $(\delta_D)_{33}^{\rm RS}$ is more than twice as large as $(\Delta_D)_{33}^{\rm RS}$, then explains why $(\Delta g_L^b)^{\rm RS}/(\Delta g_L^b)^{\rm LRS} \approx 2$ rather than the naively expected factor of about 5. In other words, non-standard contributions to the $Z^0\to b\bar b$ couplings are reduced by lowering $L$, but the effect is not as strong as anticipated in \cite{Davoudiasl:2008hx}.

\begin{figure}[!t]
\begin{center}
\vspace{3mm}
\psfrag{z}[b]{$\Lambda_{\rm UV}/\Lambda_{\rm IR}$}
\psfrag{x}[]{$L$}
\psfrag{y}[b]{$\displaystyle\frac{(\Delta g_L^b)_{\rm LRS}}{(\Delta g_L^b)_{\rm RS}}$}
\includegraphics[width=7.0cm]{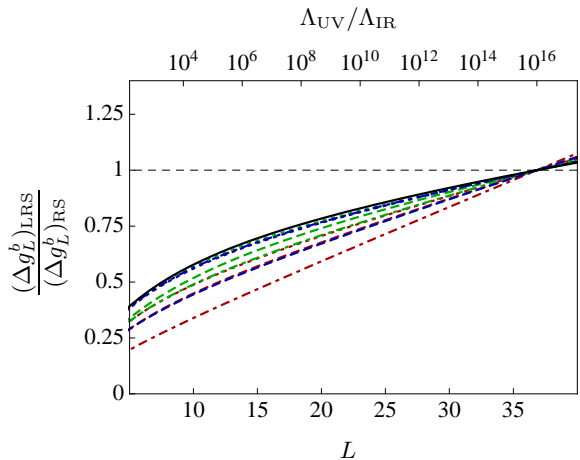}
\end{center}
\vspace{-4mm}
\caption{\label{fig:gLb} 
Ratio $(\Delta g_L^b)_{\rm LRS}/(\Delta g_L^b)_{\rm RS}$ as a function of the volume factor. The shown lines correspond to ten different sets of RS input parameters that are consistent with all relevant constraints. See text for details.}
\end{figure}

The latter feature is illustrated in Figure~\ref{fig:gLb}, which shows
the ratio $(\Delta g_L^b)_{\rm LRS}/(\Delta g_L^b)_{\rm RS}$ as a
function of the volume factor. The depicted lines correspond to ten
different sets of input parameters with fixed $\Mkk=3$\,TeV, all leading to correct predictions for the quark masses and CKM parameters as well as a consistent value of the $Z^0 b\bar b$ couplings at 95\% CL. The solid line corresponds to the default parameter set. We see that the suppression factors between $(\Delta g_L^b)_{\rm LRS}$ and $(\Delta g_L^b)_{\rm RS}$ at $L=\ln(10^3)\approx 7$ range from about 2 to 4, the exact value depending on the ratio $(\Delta_D)_{33}/(\delta_D)_{33}$ present in the original RS model. Interestingly, the contributions between $\bm{\Delta}_D$ and $\bm{\delta}_D$ can be reshuffled using the reparametrization transformation $F(c_{d_i})\to\xi^{-1}\,F(c_{d_i})$, $\bm{Y}_d\to\xi\,\bm{Y}_d$, which leaves the quark masses and mixings as well as $\bm{\Delta}_D$ invariant in the ZMA, but shifts $\bm{\delta}_D\to\xi^2\,\bm{\delta}_D$ \cite{Casagrande:2008hr}. Since this transformation involves the shift $c_{d_3}\to c_{d_3}-L^{-1}\ln\xi$, smaller values of $c_{d_3}$ accompanied by somewhat larger Yukawa couplings $|(Y_d)_{ij}|$ lead in general to an enhancement of $(\delta_D)_{33}$ over $(\Delta_D)_{33}$ and hence to a milder suppression of the curves in Figure~\ref{fig:gLb} at small $L$, similar to what is observed for our default parameter set.

To further scrutinize our generic bound $\Lambda_{\rm UV}/\Lambda_{\rm IR}>\mbox{few $10^3$}$, we have performed a detailed scan over the parameter space of the RS model for three distinct values of the volume factor, $L\in\{\ln(10^{16}),\ln(10^{5}),\ln(10^{3})\}\approx \{37,12,7\}$. In all three cases, we have randomly generated samples of 3000 points using uniform initial distributions for the input parameters. The used parameter ranges are $\Mkk\in[1,10]$\,TeV for the KK scale and $|(Y_q)_{ij}|<3$ for the Yukawa couplings. The bulk mass parameters are then chosen such that all points reproduce the observed quark masses and CKM parameters at 68\% CL. This large set of points provides a reasonable range of predictions that can be obtained for a given observable. We find that for $L=\ln(10^{16})\approx 37$ the percentage of points that satisfy the $Z^0\to b\bar b$ and $\epsilon_K$ constraints at 95\% CL amounts to 59\% and 16\%, respectively, while 13\% of the total number of generated points survive both constraints. The corresponding numbers obtained for $L=\ln(10^{5})\approx 12$ and $L=\ln(10^{3})\approx 7$ are 71\%, 24\%, 20\%, and 68\%, 11\%, 9\%. These results confirm our general findings and show that even for $\Lambda_{\rm UV}/\Lambda_{\rm IR}$ as high as $10^5$ a certain amount of tuning is required to make an anarchic RS setup consistent with the $Z^0\to b\bar b$ and $\epsilon_K$ data.

\section{Conclusions}
\label{sec:implications}

From a phenomenological point of view, the main motivation for considering volume-truncated variants of the original RS model, with a UV cutoff $\Lambda_{\rm UV}\sim 10^3$\,TeV significantly below the Planck scale, was to mitigate the strong constraints from electroweak precision tests and quark flavor physics \cite{Davoudiasl:2008hx}. We have shown that the CP-violating observable $\epsilon_K$ in $K$--$\bar K$ mixing provides a {\em lower bound\/} on the cutoff of order several $10^3$\,TeV, and that even if this bound is satisfied no significant improvement concerning $\epsilon_K$ can be achieved compared with the original RS model. The origin of this observation is the phenomenon of UV dominance of flavor-changing $\Delta F=2$ transitions, which arises whenever the bulk mass parameters determining the strange-quark mass satisfy the critical condition $c_{Q_2}+c_{d_2}<-2$. The relevant overlap integrals of the five-dimensional gluon propagator with the bulk profiles of the first- and second-generation quarks are then dominated by the region near the UV brane, thereby partially evading the RS-GIM mechanism. New-physics contributions to $\epsilon_K$ are then exponentially enhanced with respect to the original RS model addressing the hierarchy up to the Planck scale. While little RS models may still help to relax the constraint arising from the $T$ parameter, the $\Delta F=2$ flavor problem must thus be solved in another way.

\acknowledgments{One of us (MN) is grateful to the particle physics group at Melbourne University for hospitality and support during the final stage of this work, and to Tony Gherghetta for useful discussions.}

\end{document}